\documentclass[twocolumn,epj]{svjour}
\usepackage{amsmath,amssymb,bm,graphicx,tikz,cuted,widetext}
\journalname{Eur. Phys. J. C}
\usetikzlibrary{shapes}
\allowdisplaybreaks

\begin{document}
\sloppy
\title{Higgs-Higgs scattering and the (non-)existence of the Higgsonium}
\author{Vanamali Shastry\inst{a} and Francesco Giacosa\inst{a,b}}

\institute{Institute of Physics, Jan Kochanowski University, ul.  Uniwersytecka 7, P-25-406 Kielce, Poland \and Institute for Theoretical Physics, Johann Wolfgang Goethe - University, Max von Laue--Str. 1 D-60438 Frankfurt, Germany}

\date{Received: date / Revised version: date}

%\email{vanamalishastry@gmail.com}
%\email{francescogiacosa@gmail.com}

\abstract{We study the Higgs-Higgs scattering process and the possible emergence of a Higgs-Higgs bound state (Higgsonium) in any Higgs potential with the vacuum expectation value and second derivative matching the corresponding values from the Standard Model (SM). From the tree-level Higgs-Higgs scattering amplitude, we construct the unitarized amplitude using two different unitarization schemes (the well-known on-shell and N/D methods). We reproduce the known result that there is no Higgsonium state in the SM and, in addition, we determine the S-, D-, and G-wave SM scattering lengths, both at tree-level and upon unitarization. In doing so, we refine previous results by checking the convergence of the N/D approach. Next, we extend the calculation for non-SM potentials and investigate under which conditions a formation of a bound state close to the Higgs-Higgs threshold is possible. In this way, the assumption that no Higgsonium exist, imposes certain bounds on the values of the self-interaction parameters that complement those imposed by the vacuum stability condition. 
}


\maketitle

\section{Introduction}
The electroweak sector is considered apt to search for the signatures of departures from the standard model (SM). Aiding this belief are the various (but inconclusive) inconsistencies observed in the experiments. Examples include, the $R_{D^{(\ast)}}$ anomalies \cite{HFLAV:2019otj,Belle:2015qfa,MILC:2015uhg,BaBar:2013mob,Na:2015kha}, interactions of the $W^\pm$ and $Z^0$ bosons \cite{ATLAS:2022hwc,ATLAS:2022xnu}, the muon $g-2$ problem \cite{Jegerlehner:2009ry,Davier:2017zfy,Muong-2:2021vma}, etc. Most recently, the CDF collaboration reported a tension in the mass of the $W$-boson compared to the value predicted by the SM \cite{CDF:2022hxs}. Another facet of the electroweak physics is the possibility of discovering new forces of nature. The spontaneous breaking of the EW symmetry results in the $W^\pm$ and $Z^0$ bosons acquiring a large mass. This mechanism can also be implemented in scenarios where there are more than one Higgs-like state which acquire large masses upon breaking of the symmetry. Many such scenarios have been explored theoretically leading to various models \cite{Dawson:2018dcd}. The largeness of the masses of these additional Higgs-like states makes them -at present- not observable in experiments. Indirect searches for such states, however, have lead to null results \cite{Workman:2022ynf}. There is no doubt that this sector of the SM has elicited a lion's share of the interest from the theoretical as well as the experimental community. \par

A long standing topic of interest is the stability of the EW vacuum, which is related to the shape of the Higgs potential \cite{Frampton:1976kf,Linde:1977mm,Sher:1988mj,Kobakhidze:2013tn,Andreassen:2017rzq,Agrawal:2019bpm}. According to the SM, the Higgs potential is a ``Mexican hat'' potential with the Lagrangian containing a mass term, cubic term (tri-Higgs coupling, in short 3HC), and a quartic term (4HC). However, any Higgs potential that reproduces the vacuum expectation value ({\it vev}, the value of the field at the minimum of the potential, denoted as $v$) and the Higgs mass (fixed by the second derivative at the minimum and called $m_H$) is in principle acceptable, since three- or four-leg (or larger) self-interactions could not be measured yet. In this respect, the SM Higgs is the ``most economic'' choice, since there is no term beyond the four-leg one.\par

An important consequence of the uncertainty in the shape of the Higgs potential is the possibility of the vacuum decay. Simply put, if the electroweak vacuum is the global minimum of the Higgs potential, then the Universe is in a stable vacuum. Otherwise, if there are minima deeper than the ones predicted by the SM, then our Universe is not in the global minimum of the Higgs potential, and in principle, can transition into the global minimum. As such, this has been a hotly debated topic \cite{Hung:1979dn,Sher:1993mf,Casas:1996aq,Isidori:2001bm,Ellis:2009tp,Elias-Miro:2011sqh,Lebedev:2012zw,Degrassi:2012ry,Branchina:2013jra,Buttazzo:2013uya,Bednyakov:2015sca,Chigusa:2017dux,Chigusa:2018uuj}. The precise shape of the potential will depend entirely on the Higgs self-coupling constants, and understanding them is of foremost concern towards the confirmation of the SM or for finding new processes. \par
 
Many different versions of the Higgs potential have been proposed in the literature (see, {\it e.g,} Ref. \cite{Dawson:2018dcd} for a review\footnote{An important related issue of the Higgs theory is the so-called triviality, which implies that the Higgs self-interaction coupling constant(s) as well as the mass go to zero under the renormalization group (RG) evolution if the RG cut-off is taken to be infinite \cite{Maas:2012tj}. 
Due to this, the SM Higgs boson is expected to exhibit self-duality \cite{Maas:2017wzi}. The triviality problem requires that the Higgs field be associated with a finite scale. This scale (and the possible new degrees of freedom) is in itself a topic of intense debate \cite{Dawson:2018dcd}.}). These potentials include the likes of Coleman-Weinberg Higgs \cite{Hill:2014mqa}, Nambu-Goldstone Higgs \cite{Kaplan:1983fs,Kaplan:1983sm}, tadpole-induced Higgs \cite{Galloway:2013dma,Chang:2014ida}, etc which behave like the SM Higgs potential only in the vicinity of the SM minima, and the extensions of the SM - like the additional singlet \cite{Silveira:1985rk,OConnell:2006rsp,Bowen:2007ia,Espinosa:2011ax,Pruna:2013bma}, two Higgs doublets (2HDM) \cite{Haber:1978jt,Gunion:2002zf,Branco:2011iw}, minimally supersymmetric SM Higgs (MSSM) \cite{Gunion:1984yn,Gunion:1986nh}, etc. Another common feature of all these potentials is that the {\it vev} of the field (in case of doublet models allowing more than one {\it vev}'s, then the sum of the squares of the {\it vev}'s, see {\it e.g.}, Ref. \cite{Dawson:2018dcd} for exceptions to this rule) agrees with that of the SM. Furthermore, in models with more than one Higgs-like field, the physical Higgs either emerges as an admixture of the fields \cite{Lopez-Val:2013yba,Chen:2014ask} or is decoupled from the rest (see {\it e.g,} Ref. \cite{Carena:2013ooa,BhupalDev:2014bir}). In the former case, the mixing angle(s) can be constrained using the mass of the physical Higgs. However, the Higgs self-coupling are (largely) unconstrained. It is pertinent to note that searches for additional Higgs have so far provided null results \cite{ATLAS:2015pre,ATLAS:2017eiz,CMS:2018rmh}.\par

The experimental data on the Higgs self-couplings are scarce. This is partly due to the fact that the current state-of-art is insufficient to perform Higgs-Higgs scattering experiment, and the accessible probes like di-Higgs and tri-Higgs production through gluon-gluon fusion, the top quark-Higgs ($t\bar{t}hh$) coupling, and so on have large backgrounds as well as non-perturbative effects arising from strong interactions that limit the accuracy of the measurements \cite{Baur:2003gp,Dolan:2012rv,Goertz:2013kp}. 
The possibility of measurements of the self-couplings at various future experiments has been studied in various works \cite{Dolan:2012rv,Goertz:2013kp,Baglio:2012np,McCullough:2013rea,Cao:2015oxx,Cao:2016zob,DiVita:2017eyz,Degrassi:2021uik,Chiesa:2020awd,Park:2020yps,Alasfar:2022zyr} and is a topic of great interest for the community \cite{Dawson:2022zbb,Abu-Ajamieh:2022dtm,Apresyan:2022tqw}. Two alternate avenues - pair production of the Higgs \cite{ATLAS:2019qdc,ATLAS:2021ifb,ATLAS:2022xzm,ATLAS:2023qzf} and the single Higgs production process \cite{Degrassi:2016wml,Maltoni:2017ims,ATLAS:2022vkf} - were explored to study the self-coupling of the Higgs. The tri-Higgs coupling constant was extracted from the resonant and non-resonant pair production of the Higgs boson in the $b\bar{b}$ channels, as well as the $pp\to HH$ channel. The resonant production of Higgs studied by ATLAS provided the bounds of $-5 < d_3 <12$ \cite{ATLAS:2019qdc} which was later revised to $-0.4<d_3<6.3$ using additional data and by including the single Higgs production channels \cite{ATLAS:2022jtk}. The non-resonant production studied by CMS provided the estimate $-1.24 <d_3 <6.49$ \cite{CMS:2022dwd}.
Hopefully, the knowledge about this part of the SM can be improved and resolved in more detail in future colliders \cite{Homiller:2018dgu,Cepeda:2019klc,Millet:2019oqe,Black:2022nzv,Taliercio:2022maa,Bagger:2022zyy}. \par 

Moreover, if the extended Higgs potential has a suitable form, with the 3HC and 4HC coupling constants taking suitable values, one can expect to have a di-Higgs bound state or a {\it Higgsonium} (also called the {\it Higgsium}). 
A study of the relationship between the coupling constants and the Higgsonium can shed light on the shape of the potential around the minima. One of the earliest studies of the formation of the bound state of two Higgs was performed by Cahn and Suzuki using a model similar to the linear sigma model \cite{Cahn:1983vi}. This model is essentially the same as the non-interacting Higgs doublet model. They find that a (shallow) two-Higgs bound state could exist only if the Higgs has a mass $>1$ TeV (which is presently excluded). Similar results were obtained with interacting Higgs using the $N/D$ unitarization scheme \cite{Contogouris:1988rd} and with the SM Higgs using Bethe-Salpeter equations \cite{Rupp:1991bb}. An analogous outcome was found through a relativistic calculation using the minimal SM Lagrangian \cite{DiLeo:1994xc}. An extension of that work to the SM Lagrangian showed that a bound state can be observed for a light Higgs (in the vicinity of the physical mass of the Higgs boson), only if the Higgs self-coupling constants are large \cite{Siringo:2000kh}. A nonrelativistic study of interactions between two Higgs in a two-Higgs doublet model also showed the presence of Higgsonium even when the mass of Higgs is smaller than the presently known value \cite{Grifols:1991gw,Clua:1995ni}.\par

A systematic study based on the linear realization of the Higgs effective field theory (HEFT) showed that a Higgsonium is possible in the strong regime of Higgs self-coupling, reaffirming the earlier results \cite{Grinstein:2007iv}. The nonrelativistic HEFT derived in Ref. \cite{Grinstein:2007iv} was extended to the two-Higgs doublet model (2HDM) in Ref. \cite{Biswas:2019arj}, that showed that the heavy partners of the Higgs - called the ``obese Higgs'' - can form bound states in the Type-I 2HDMs and not in the type-II 2HDMs, provided the mass of the ``obese Higgs'' is less than the scale of expansion of the HEFT (and greater than the mass of the SM Higgs). That work also explored the decays of such a bound state into various channels. \par
\begin{figure*}[ht]
\centering
\begin{tikzpicture}
\draw (0,0) -- (3,2);
\draw (0,2) -- (3,0);
\filldraw[red] (1.4,1.1) rectangle (1.6,0.9);
\node at (1.5,-0.5) {(a)};

\draw (4,0) -- (5,1) -- (4,2);
\draw (5,1) -- (6,1);
\filldraw[black] (5,1) circle (2pt);
\draw ((7,0) -- (6,1) -- (7,2);
\filldraw[black] (6,1) circle (2pt);
\node at (5.5,-0.5) {(b1)};

\draw (8,2) -- (9,1.5) -- (10,2);
\draw (9,1.5) -- (9,0.5);
\filldraw[black] (9,1.5) circle (2pt);
\draw (8,0) -- (9,0.5) -- (10,0);
\filldraw[black] (9,0.5) circle (2pt);
\node at (9,-0.5) {(b2)};

\draw (11,2) -- (12,1.5) -- (13,0);
\draw (12,1.5) -- (12,0.5);
\filldraw[black] (12,1.5) circle (2pt);
\draw (11,0) -- (12,0.5) -- (13,2);
\filldraw[black] (12,0.5) circle (2pt);
\node at (12,-0.5) {(b3)};
\end{tikzpicture}
\caption{Tree level diagrams representing the $HH\to HH$ scattering process. The red square and black circles represent the vertices coming from the $H^4\text{ and }H^3$ terms respectively.}\label{treescatdia}
\end{figure*}
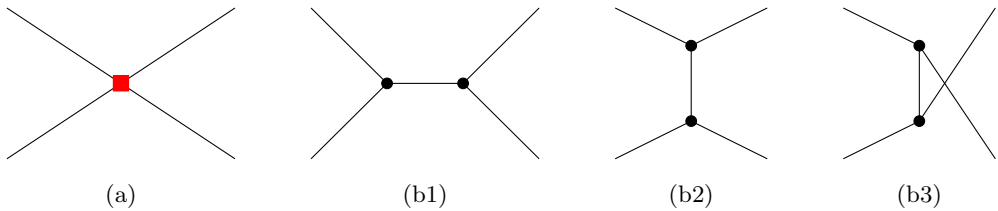
This paper is organized as follows. In Sec. \ref{sec:formal} we detail the formalism used in the study. We present the results of the calculation in Sec. \ref{sec:RnD} and conclude in Sec. \ref{sec:SnC}.\par

In the present work, we are concerned with a related possible avenue for the studying the Higgs self-coupling.
Namely, we are interested in the study of the Higgs-Higgs scattering and the consequent determination of the (S-, D-, and G-wave) scattering lengths and phase shifts in both the SM Higgs potential and in extensions of the latter. 
An important part of this work concerns the unitarization of the scattering results of the SM Higgs Lagrangian (and extensions thereof). To this end, we use the well-known on-shell and the already mentioned $N/D$ schemes to improve the tree-level results for Higgs-Higgs scattering (scattering length and phase shift). The use of two schemes allows for a direct comparison, thus checking the independence of the result on the adopted unitarization. In particular, for the $N/D$ case, we also go beyond the usually employed lowest order approximation and show (to our knowledge for the first time) the convergence of the results. 
Moreover, we also revisit the possibility of the formation of the bound state of two Higgs bosons by investigating in a systematic way the whole parameter space for the three- and four-leg interaction strengths, upon keeping the {\it vev} ($v$) and the mass ($m_H$) fixed at their physical values. We find that there is a minimum value of the 3HC coupling constant {\it below} which no bound state can exist and a lower bound for the 4HC coupling constant {\it above} which the repulsion does not allow for a bound state. In this way, assuming that no Higgsonium exists, a further constrain of the 3H-4H parameter space that improves previously known bounds is determined.\par

Quite interestingly, the study of scattering and the possible emergence of bound states has been recently also tackled in other areas of physics: in Ref. \cite{Giacosa:2021brl} the bound states of two dilatons/glueballs has been studied, finding that it is quite possible to have such a state in YM theory of QCD. This is particularly interesting since the dilaton potential resembles the Higgs one (indeed, it is similar to the Coleman-Weinberg Higgs mentioned above \cite{Hill:2014mqa}). Bound states involving glueballs have been later on investigated in Ref. \cite{Petrov:2022rkn}. An interesting novel development in the study of bound states has been recently applied to the scattering of gravitons \cite{Guerrieri:2021ivu,Blas:2020och}, leading to the graviball hypothesis in Ref. \cite{Blas:2020och}.\par
 
\section{The Higgs potential\label{sec:formal}}
\subsection{General considerations}
Consider a potential $V(H)$ which has the basic features of the Mexican hat potential in the relevant part of the domain, but can vary arbitrarily at large values of the field or when the field is close to zero. The potential must,
\begin{enumerate}
    \item possess minima at $H=\pm v$, where $v \sim 246$ GeV is the {\it vev} \cite{Workman:2022ynf}, and
    \item have a value equal to $m_H$ for the second derivative at the minima ($H=\pm v$).
\end{enumerate}
If the potential is well behaved around $H=\pm v$, we can expand it in a Taylor series as,
\begin{align}
   V(H) &= V(v) + \frac{m_H^2}{2!} (H-v)^2 + \frac{g}{3!} (H-v)^3 + \frac{\lambda}{4!} (H-v)^4\nonumber\\
   & + \frac{g_{5H}}{5!} (H-v)^5 + \ldots\label{sympot2}\\
   &= V(v)+ \frac{m_H^2}{2!} h^2 + \frac{g}{3!} h^3 + \frac{\lambda}{4!} h^4 + \frac{g_{5H}}{5!} h^5 + \ldots\label{sympot3}
\end{align}
where $\left(dV/dH\right)|_{H=v} = 0$ because of the extremum. In the last line, $h = H - v$ has been defined. Note that the coupling constants $g_{nH}$ represent $n-$Higgs interactions, with $g_{2H}=m_H^2$, $g_{3H}=g$, and $g_{4H}=\lambda$.\par

The assumptions listed above restrict the mass of the Higgs boson and the $vev$ of the Higgs potential. However, the coupling constants $g$ and $\lambda$ remain unrestricted. Thus, the current experimental data can fix only the position of the minimum of the Higgs field and its second derivative around the minimum. The shape of the potential in general is not probed. We use this observation to express the Higgs self-coupling constants as in Ref. \cite{Agrawal:2019bpm},
\begin{align}
g = d_3 \frac{3 m_H^2}{v} &\quad \lambda = d_4 \frac{3 m_H^2}{v^2}
\end{align}
The numerical values $d_3=d_4=1$ correspond to the ``standard model" self-coupling and any deviation from these values would hint a disagreement with it. In the discussions below, we alternate between $(g,\lambda)$ and $(d_3,d_4)$ based on convenience. (This is also true for the notations $h$ and $H$ representing the Higgs field.)\par
\begin{figure*}[ht]
\centering
\includegraphics[scale=0.65]{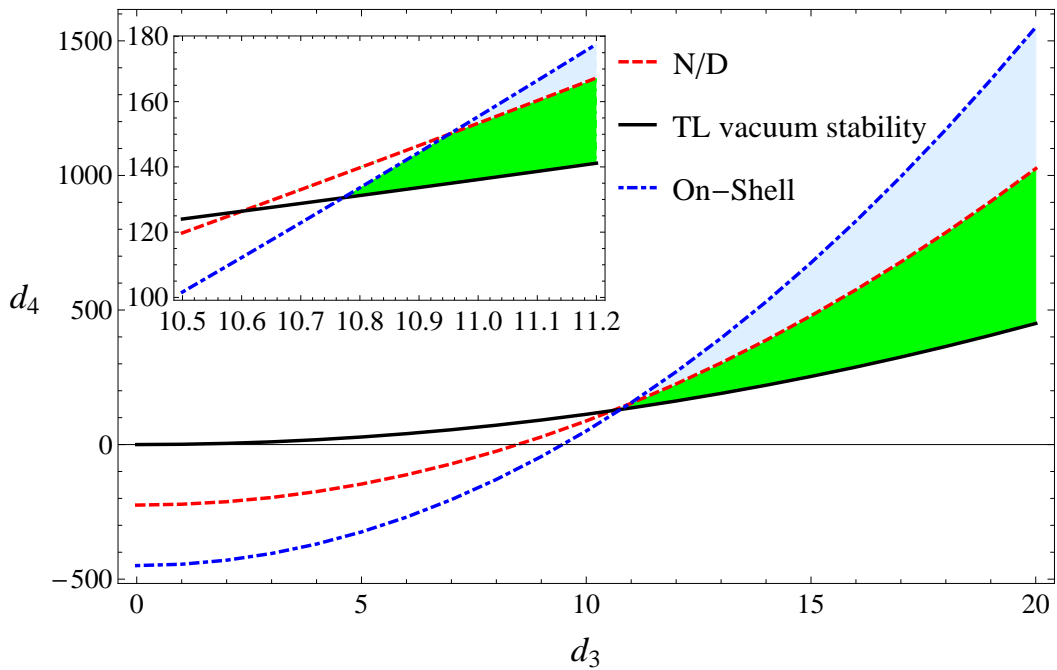}
\caption{Assuming the non-existence of the Higgsonium, the allowed parameter space in the $({d_3,d_4})$-plane lies above the dotted-dashed (on-shell) and the dashed ($N/D$) curves. The requirement that the vacuum of the Higgs potential corresponds to an absolute minimum implies the region above the solid line \cite{Agrawal:2019bpm}. As it is visible, the non-existence of the Higgsonium further constraints the parameter space in the right-upper part of the $({d_3,d_4})$-plane. Inset: The region where the different curves cross each other (see text for details).}\label{fig:bounds}
\end{figure*}

\subsection{Higgs-Higgs scattering}
At tree-level, only the cubic and quartic terms contribute to the $HH\to HH$ scattering, as shown in the Fig. \ref{treescatdia}. The amplitudes for these diagrams are given by,
\begin{align}
    i\mathcal{M}_a &= -i\lambda\\
    i\mathcal{M}_{b1} &= -i g^2 \frac{1}{s-m_H^2 + i\epsilon}\\
    i\mathcal{M}_{b2} &= -i g^2 \frac{1}{t-m_H^2 + i\epsilon}\\
    i\mathcal{M}_{b3} &= -i g^2 \frac{1}{u-m_H^2 + i\epsilon}
\end{align}
where $s,t,u$ are the Mandelstam variables. Upon projecting the amplitudes to the $s$-channel and later onto various angular momentum channels, we get the partial wave amplitudes as,
\begin{align}
    A_0(s) &= -\lambda-\frac{g^2}{s-m_H^2}+\frac{2 g^2 \log \left(\frac{s-4
   m_H^2}{m_H^2}+1\right)}{s-4 m_H^2} \text{ ,}\label{eq:scatamp}\\
   A_{\ell\ge 2}(s) &= \frac{g^2}{k^2}Q_\ell\left(1+\frac{m_H^2}{2k^2}\right)\text{ ,} \label{eq:scatampHW}
\end{align}
where $\ell$ is the orbital angular momentum quantum number, $Q_\ell(s)$ are the Legendre functions, and $k=\frac{1}{2}\sqrt{s-4m_H^2}$ is the 3-momentum of any of the final states in the center of mass frame of reference. The odd-$\ell$ partial waves vanish due to bosonic nature of the Higgs. One crucial observation is that the $\ell=0$ partial wave contains a branch point at $s=3m_H^2$ in addition to the pole at $s=m_H^2$. This feature is used in the so-called ``on-shell'' unitarization scheme followed in the present work.\par

Finally, the $s$-wave tree-level scattering length ($a_\ell$) reads\footnote{ We follow the convention where $\delta_0 = k~\!a_0$ \cite{Broniowski:2015oha}. Thus, our scattering length varies by a sign compared to that of \cite{Feng:2013jza}.}
\begin{align}
a^{TL}_0 &= \dfrac{1}{32\pi m_H}A_0(s=4m_H^2) = \frac{1}{32\pi m_H}\left(-\lambda +\frac{5g^2}{3m_H^2}\right)\nonumber\\
   &= \dfrac{3m_H}{32\pi v^2} (5d_3^2-d_4) 
   \text{ .}
   \label{a0tleq}
\end{align}
For the SM parameters, we get:
\begin{equation}
   a^{TL}_0= \dfrac{3m_H}{8\pi v^2}  = (4.86\pm 0.01)\times10^{-5} \text{ fm} \label{a0tl}
\end{equation}
which is in good agreement with the result from nonrelativistic effective theory calculations \cite{Feng:2013jza}. Such a small value for the $s$-wave scattering length is a sign of the weakness of the interaction between the Higgs states. For comparison, the $s$-wave scattering length for $W^\pm W^\mp$, and $W^\pm Z^0$ scattering is nearly 3 orders of magnitude higher\footnote{ It should be noted that the scattering between the $W^\pm$ and/or $Z^0$ states are inelastic and hence the scattering lengths have complex values. Thus, in the comparison above one refers to their absolute values.} \cite{Butterworth:2002tt,Ballestrero:2011pe,Chang:2013aya}.
The smallness of the SM $HH$-scattering length shall be confirmed later on by the very small unitarity corrections. For the higher partial waves, we can evaluate the scattering lengths (or better, hyper-volumes). The general expressions for the scattering length
is,
\begin{align}
    a_{\ell} &= \frac{1}{32\pi m_H} \lim_{s\to 4m_H^2} \dfrac{1}{k^{2\ell}}A_\ell(s) \text{ .}
\end{align}
Using Eq. \ref{eq:scatampHW}, we get the values for $\ell=2$ and $\ell=4$ partial waves as,
\begin{align}
    a^{TL}_2 &= \dfrac{g^2}{30\pi m_H^7} = \dfrac{3d_3^2}{10\pi v^2 m_H^3} \nonumber\\
   & \stackrel{\text{SM}}{=} \dfrac{3}{10\pi v^2 m_H^3} = 2.4 \times 10^{-16} \text{ fm}^{5} \text{ ,}\\
    a^{TL}_4 &= \dfrac{8g^2}{315\pi m_H^{11}} = \dfrac{8d_3^2}{35\pi v^2 m_H^7}\nonumber\\
   & \stackrel{\text{SM}}{=} \dfrac{8}{35\pi v^2 m_H^7} = 1.1\times 10^{-27}\text{ fm}^9 \text{ .}
\end{align}
Interestingly, the scattering amplitudes in higher partial wave channels involve only the attractive three-leg interaction. Later on, we shall reevaluate the scattering within the introduced unitarization schemes. Remarkably, for the SM case, the modification turn out to be completely negligible.\par

In the future, it would be also interesting to calculate how additional SM processes (as for instance the box diagrams with top quarks) would modify, within the SM, the effective parameters $d_3$ and $d_4$. Due to the suppression of the involved diagrams, one should expect only small deviations from $d_3=d_4=1$. Note that the effects of the quantum corrections on the Higgs self-coupling constants have been studied up to the next-to-next-to-leading order in various extensions of the SM. 

The estimates vary depending on the type of extensions considered as well as the parameters of the respective models \cite{Kanemura:2004mg,Senaha:2018xek,Braathen:2019zoh,Braathen:2019pxr} but is of (maximally) order $10^2$ for $d_3$.
Thus, the qualitative conclusions arrived at in the present work remain valid and Eq. \ref{a0tleq} could be still used to determine the scattering length for the updated `dressed' values of $d_3$ and $d_4$.

\begin{figure*}[ht]
\centering
\includegraphics[scale=0.65]{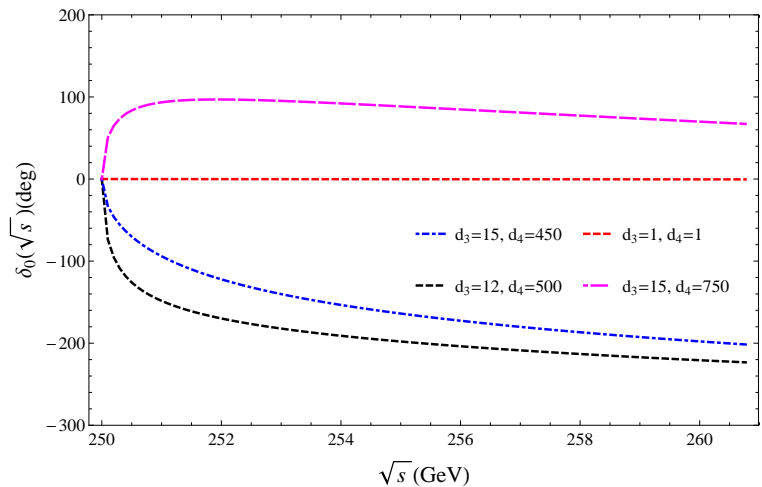}
\caption{Plots of the $S$-wave scattering phase shift for various values of the parameters.}\label{fig:phshift}
\end{figure*}

\section{Unitarization\label{sec:unit}}
In this section we discuss the main features of a generic unitarization scheme and present two concrete ones: the on-shell and the $N/D$ schemes. Then, we shall apply them to the Higgs-Higgs scattering process. 
\subsection{General considerations}
The amplitude calculated in the previous section was obtained at tree-level,
which, as it is known, fulfills unitarity only order-by-order in perturbation theory.
An effective way to respect unitarity (as well as including the effect of quantum fluctuations) is possible via an appropriate unitarization scheme that provides a prescription to replace the amplitude
\begin{equation}
A_{l}(s)\rightarrow A_{l}^{U}(s) \text{ ,}
\end{equation}%
where $A_{l}^{U}(s)$ is the unitarized amplitude. A unitarized amplitude typically consists of (some) contributions coming from all orders of perturbation involving the vertices and degrees of freedom present in the tree-level amplitude. 
In the literature, different unitarization schemes have been investigated (see e.g. Refs. 
\cite{Giacosa:2021brl,Dobado:1992ha,Gulmez:2016scm,Oller:1997ng,Oller:2020guq,Mai:2022eur,Black:2000qq,Guo:2006br,Nieves:1998hp,Nieves:2001de,Salas-Bernardez:2020hua,Delgado:2015kxa,Pelaez:2015qba} and refs therein). These schemes have been used to study a wide variety of topics including, but not limited to $\pi\pi$, $\pi K$ \cite{Dobado:1992ha}, $\rho\rho$ scattering \cite{Gulmez:2016scm}, and lately to the scattering of two scalar glueballs and the possibilities of a ``glueballonium" \cite{Giacosa:2021brl},etc. 
In fact, the unitarization is a common tool in the theoretical search for short-lived (a.k.a, broad) resonances like the scalar mesons \cite{Pelaez:2015qba}.\par

Before discussing the two chosen approaches (on-shell and $N/D$) in more in detail,
we provide some relevant general properties of a scattering process.
For simplicity, we concentrate on the S-wave only $(\ell=0)$, whose index is
omitted. The phase-shift $\delta(s)$ is linked to the unitarized
amplitude $A^{U}(s)$ by the following relation:%
\begin{equation}
\frac{e^{2i\delta (s)}-1}{2i}=\rho (s)A^{U}(s)\text{ ,}  \label{ps}
\end{equation}%
where the phase-space function $\rho (s)$ reads%
\begin{equation}
\rho (s)=\frac{1}{2}\frac{\sqrt{\frac{s}{4}-m_{H}^{2}}}{8\pi \sqrt{s}}\text{
.}
\end{equation}%
Since for elastic scattering $\delta (s)$ is purely real, Eq. \ref{ps}
implies that:%
\begin{equation}
\text{Im}\left( A^{U}(s)^{-1}\right) =-\rho (s)\text{ for }s>4m_{H}^{2}\text{
.}  \label{con}
\end{equation}%
This is a necessary condition for any unitarization approach. \par

The emergence of the bound state is also connected to the value of the unitarized scattering length, which, for the S-wave, reads
\begin{equation}
a^{U} = \frac{1}{32\pi m_H}A^{U}_{\ell=0}(s=4m_H^2) \text{ .}
\end{equation}
When a bound state forms right at threshold, the scattering length diverges. Moreover, the scattering length is positive when no bound state forms and negative otherwise. 

\begin{figure*}[ht]
    \centering
    \includegraphics[scale=0.67]{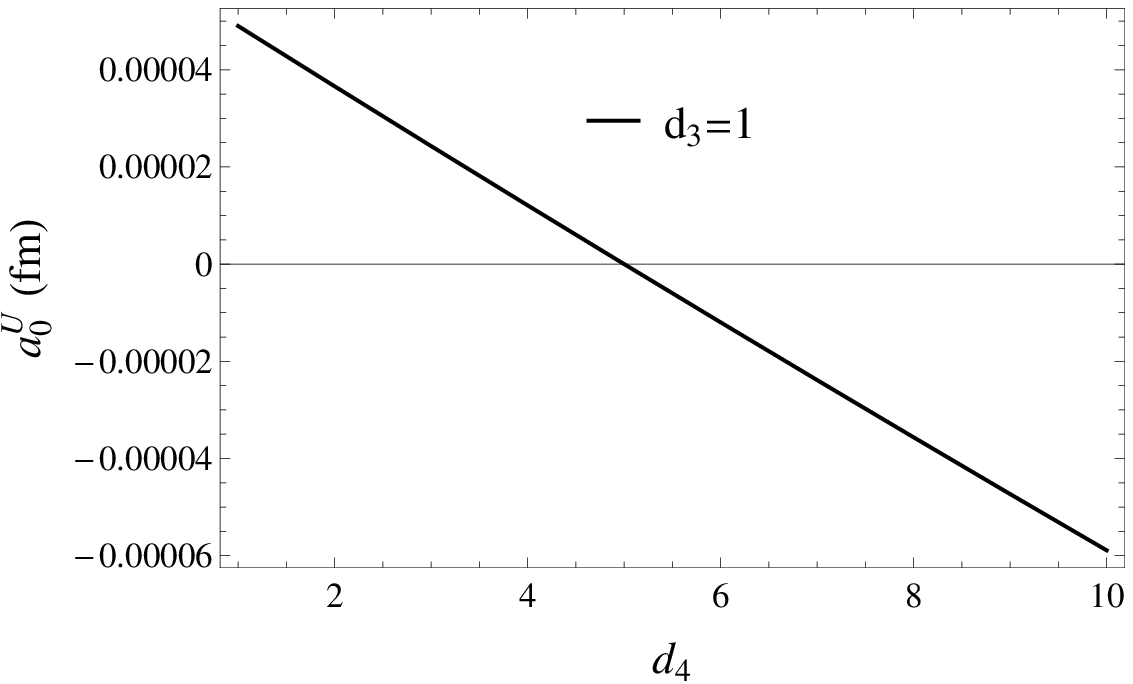}~\includegraphics[scale=0.65]{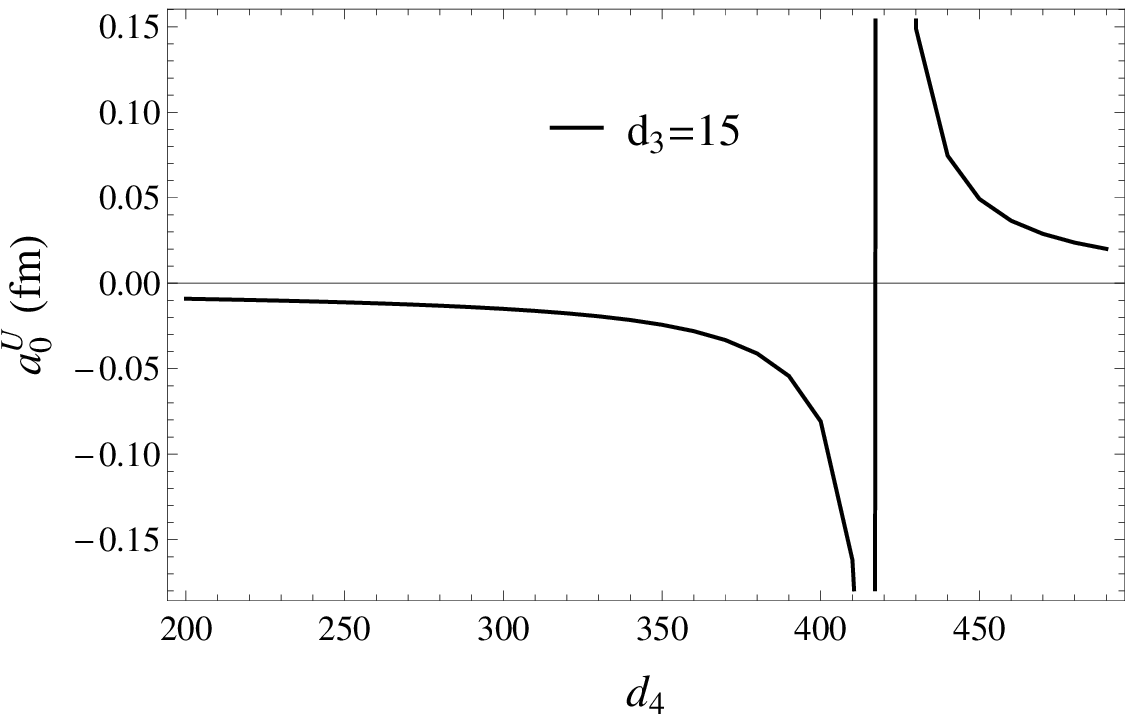}
    \caption{The Higgs-Higgs scattering length as function of $d_4$ for two different values of $d_3$. The starting point on the left graph gives the value for the SM parameters. Numerically, it is very close to the tree-level value of Eq. \ref{a0tl}}.
    \label{fig:scatlen}
\end{figure*}

\subsection{On-shell unitarization}
In the on-shell unitarization approach \cite{Dobado:1992ha,Oller:1997ng,Samanta:2020pez,Samanta:2021vgt},
one introduces the loop function $\Sigma (s)$ entering into the resummation 
\begin{equation}
A^{U}(s)=A(s)+A(s)\Sigma (s)A(s)+...=\left[ A^{-1}(s)-\Sigma (s)\right] ^{-1}%
\text{,}
\end{equation}%
where $\text{Im}\left(\Sigma (s)\right)=\rho (s)$ for $s>4m_{H}^{2}$ is required to
fulfill Eq. \ref{con}.
The main feature of this approach is that the
tree-level amplitude and the loop factorize.

The loop function is fixed by considering two subtractions \cite%
{Giacosa:2021brl}: 
\begin{align}
\Sigma (s)&=-\dfrac{(s-m_{H}^{2})(s-3m_{H}^{2})}{\pi }\nonumber\\
   &\int_{4m_{H}^{2}}^{%
\infty }\frac{\rho (s^{\prime })}{(s-s^{\prime }+i\varepsilon )(s^{\prime
}-m_{H}^{2})(s^{\prime }-3m_{H}^{2})}ds^{\prime }\text{ ,}
\end{align}%
whose analytical form is:
\begin{align}
    \Sigma(s) &= \frac{1}{192\sqrt{3}\pi m_H^2}(7m_H^2-s)\nonumber\\
   &+\frac{k}{16\pi^2\sqrt{s}}\log\left(\frac{\sqrt{4m_H^2-s}+i\sqrt{s}}{\sqrt{4m_H^2-s}-i\sqrt{s}} \right) \text{ .}
\end{align}

Note, while a single subtraction is enough for convergence, two subtractions
are requested for avoiding the emergence of unphysical ghost states \cite{Donoghue:2019fcb,Trotti:2022ukp}. 
Here, the first subtraction is taken at $s=m_{H}^{2}$ in such a way to
maintain the Higgs mass at the same physical value even after performing the
unitarization. The second subtraction is chosen at $s=3m_{H}^{2}$, in order to preserve the
divergence of the tree-level amplitude at the branch point of the left-hand
cut. In fact, this branch point is also caused by the single particle pole
at $t=m_{H}^{2}$ and $u=m_{H}^{2}$ projected onto the $s$-channel. In other words, for $s \sim 3m_H^2$, the unitarized S-wave amplitude is given by $A^{U}(s) \sim A(s) \sim -2(g/m_H)^2 \log((s-3m_H^2)/m_H^2)$.\par

Finally, a bound state with mass $M_{2H}$ is a solution of the equation 
\begin{equation}
1-A(s=M_{2H}^{2})\Sigma (s=M_{2H}^{2})=0 \text{ ,}
\label{th}
\end{equation}%
that implies that the S-wave scattering length diverges when $M_{2H} = 2m_H$. This is due to the fact that a bound state just at threshold generates a divergent pole of the amplitude.
A feature of the on-shell method is that $M_{2H}\in (\sqrt{3}m_{H},2m_{H}),$ but the
lower limit $\sqrt{3}m_{H}$ should not be regarded as a physical constraint. In turn,
it implies that this unitarization scheme should not be trusted when $M_{2H}$ is far below the threshold.\par

The unitarized scattering length in the on-shell approach reads
\begin{align}
a^{U} &= \frac{1}{32\pi m_H}\frac{-\lambda +\frac{5g^2}{3m_H^2}}{1-\left(-\lambda +\frac{5g^2}{3m_H^2}\right)\Sigma(s=4m_H^2)} 
 \text{ ,}
\end{align}
where 
\begin{equation}
\Sigma(s=4m_H^2)=\frac{3}{192 \sqrt{3} \pi}
\end{equation}
is the total contribution of the loops to the scattering length. For the SM parameters ($d_3=d_4=1$), this value is extremely small compared to the inverse of the tree-level amplitude at threshold, $A(s=4m_H^2)$ and hence can be neglected. Thus, the value of the unitarized scattering length is nearly the same as that obtained at the tree-level. This conclusion applies with even better accuracy to the higher D- and G-waves, as explicitly shown in Ref. \cite{Giacosa:2021brl}.\par

The irrelevance of the unitarization (and thus of quantum fluctuations) does not, however, apply in general: the closer one gets to the generation of a bound state, the more important are the loops.\par
\begin{figure*}[ht]
\centering
\includegraphics[scale=0.65]{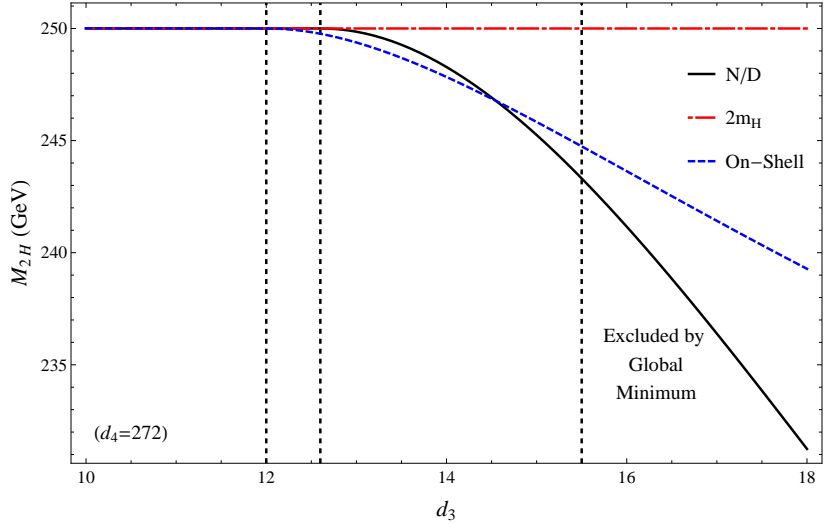}
\caption{Plot of the mass of the Higgsonium as a function of $d_3$ for a representative value of $d_4(=272)$. The black solid line represents the value from the N/D unitarization and the blue dashed line gives the value from the on-shell unitarization. The first two vertical lines represent (from left to right) the critical values for having an Higgsonioum in the on-shell and $N/D$ schemes, and the third vertical line marks the region excluded by the global-minimum requirement. }\label{fig:2hmass}
\end{figure*}

\subsection{The $N/D$ unitarization}
In the $N/D$ approach \cite%
{Frazer:1969euo,Hayashi:1967bjx,Gulmez:2016scm,Oller:2020guq,Cahn:1983vi,Mai:2022eur,Giacosa:2021brl},
the unitarized S-wave amplitude takes also into account the left-hand cut
behavior. Namely, at tree-level, one has 
\begin{equation}
\text{Im}(A)=\sigma(s) =\frac{2g^{2}}{s-4m_{H}^{2}} \text{ for } s\leq 3m_{H}^{2} 
\text{ .}
\label{sigma}
\end{equation}%
This is due to the single-particle pole at $t=u=m_{H}^{2}.$ 
Here,
Eq. \ref{sigma} is fulfilled by requiring that the unitarized amplitude is
expressed as a ratio 
\begin{equation}
A^{U=\text{N/D}}(s)=\frac{N(s)}{D(s)} \text{ ,}
\end{equation}%
where the numerator $N$ contains the left-hand cut and the denominator $D$
the right-hand cut:%
\begin{equation}
\text{Im}(N)=\left\{ 
\begin{array}{c}
\sigma (s)D(s)\text{ for }s\leq 3m_{H}^{2} \\ 
0\text{ for }s>3m_{H}^{2}%
\end{array}%
\right.
\text{ ,}
\label{num}
\end{equation}%
and
\begin{equation}
\text{Im}(D)=\left\{ \text{ }%
\begin{array}{c}
0\text{ for }s<4m_{H}^{2} \\ 
-\rho (s)N(s)\text{ for }s\geq 4m_{H}^{2}%
\end{array}%
\right.
\text{ .}
\label{den}
\end{equation}%
By construction, Eq. \ref{con} and Eq. \ref{sigma} are used. We require also that $D(m_{H}^{2})=1$, thus the
residue of the amplitude at the single-particle pole is unchanged by quantum fluctuations.\par

The previous constraints lead to the following coupled integral equations for 
$N$ and $D$:%
\begin{align}
N(s) &=A(s)+\dfrac{1}{\pi }\int_{-\infty }^{3m_{H}^{2}}ds_{2}\frac{\sigma
(s_{2})\left( D(s_{2})-1\right) }{(s_{2}-s-i\varepsilon )} \text{ ,} \\
D(s) &=1-\dfrac{(s-m_{H}^{2})}{\pi }\int_{4m_{H}^{2}}^{\infty }\frac{\rho
(s_{1})N(s_{1})}{(s_{1}-s-i\varepsilon )(s_{1}-m_{H}^{2})}ds_{1\text{ }},
\end{align}%
where we have taken into account that the numerator reduces to the
tree-level amplitude at lowest order (see below). The mass of the Higgsonium bound
state is obtained by 
\begin{equation}
D(s=M_{2H}^{2})=0\text{ .}
\end{equation}

A useful method to solve the system above is by iterations. By denoting $%
N^{(k)}(s)$ and $D^{(k)}(s)$ as the functions $N$ and $D$ at the $k^\text{th}$
iteration, at lowest order we set: 
\begin{equation}
N^{(0)}(s)=A(s)\text{ , }D^{(0)}(s)=1\text{ ,}
\end{equation}%
thus the amplitude $A^{N/D(0)}(s)=N^{(0)}(s)/D^{(0)}(s)=A(s)$ (tree-level
result). At the next order, the denominator takes the form 
\begin{equation}
D^{(1)}(s)=1-\dfrac{(s-m_{H}^{2})}{\pi }\int_{4m_{H}^{2}}^{\infty }\frac{%
\rho (s_{1})N^{(0)}(s_{1})}{(s_{1}-s-i\varepsilon )(s_{1}-m_{H}^{2})}ds_{1}\label{eq:D1approx}
\end{equation}%
which leads to the first non-trivial $N/D$ result \cite{Cahn:1983vi,Gulmez:2016scm} with 
\begin{align}
A^{N/D(1)}(s)&=\frac{N^{(0)}(s)}{D^{(1)}(s)} \text{ ,}
\end{align}%
where $D^{(1)}(s)$ fulfills $\text{Im}\left(D^{(1)}(s)\right)=-\rho (s)N^{(0)}(s),$
while $\text{Im}\left(N^{(0)}\right)=\sigma (s)$. Note, the single subtraction in the
denominator is taken at $s=m_{H}^{2}$ giving $D^{(1)}(m_{H}^{2})=1$, as
desired. Here, there is no need for further subtractions, since no ghost appears in this
approach. Moreover, the $N/D$ amplitude automatically keeps the branch point $A_0^{U = N/D}(s) \propto \log((s-3m_H^2)/m_H^2)$ (even though the constant in front of it is modified by the unitarization).

At this order, the bound state equation is $D^{(1)}(s=M_{2H}^{2})=0.$ Quite
remarkably, it turns out that the results are quite accurate, since higher
orders cause only small changes, as we show in Appendix A. Due to its
relatively simple implementation, we shall use this approximation in the
following plots.

Going further, one needs to evaluate the chain:%
\begin{equation}
A = N^{(0)}\rightarrow D^{(1)}\rightarrow N^{(1)}\rightarrow
D^{(2)}\rightarrow \ldots
\end{equation}%
leading to the function $N^{(1)}(s)$ and $D^{(2)}(s)$ given by: 
\begin{align}
N^{(1)}(s)& =A_{0}(s)+\dfrac{1}{\pi }\int_{-\infty }^{3m_{H}^{2}}ds_{2}\frac{%
\sigma (s_{2})\left( D^{(1)}(s_{2})-1\right) }{(s_{2}-s-i\varepsilon )}\text{
;} \\
D^{(2)}(s)& =1-\dfrac{(s-m_{H}^{2})}{\pi }\int_{4m_{H}^{2}}^{\infty }\frac{%
\rho (s_{1})N^{(1)}(s)}{(s_{1}-s-i\varepsilon )(s_{1}-m_{H}^{2})}ds_{1}\text{
.}
\end{align}%
At an arbitrary iteration:
\begin{align}
N^{(k)}(s)& =A_{0}(s)+\dfrac{1}{\pi }\int_{-\infty }^{3m_{H}^{2}}ds_{2}\frac{%
\sigma (s_{2})\left( D^{(k)}(s_{2})-1\right) }{(s_{2}-s-i\varepsilon )}\text{
;} \label{eq:Nk}\\
D^{(k+1)}(s)& =1\nonumber\\
   &-\dfrac{(s-m_{H}^{2})}{\pi }\int_{4m_{H}^{2}}^{\infty }\frac{%
\rho (s_{1})N^{(k)}(s)}{(s_{1}-s-i\varepsilon )(s_{1}-m_{H}^{2})}ds_{1}\text{
,}\label{eq:Dkplus1}
\end{align}%
with $\text{Im}\left(D^{(k+1)}(s)\right)=-\rho (s)N^{(k)}(s)$ and $\text{Im}%
\left(N^{(k)}(s)\right)=\sigma (s)D^{(k)}(s).$ Finally, the iterative $N/D$ amplitudes
read:%
\begin{equation}
A^{N/D(2k)}(s)=\frac{N^{(k)}(s)}{D^{(k)}(s)}\text{ , }A^{N/D(2k+1)}(s)=\frac{%
N^{(k)}(s)}{D^{(k+1)}(s)}
\text{ .}
\end{equation}%
Clearly, the bound states are searched as poles of \\$A^{N/D(2k+1)}(s),$ alias the
zeroes of $D^{(k+1)}$ need to be determined.
Within the present method, the mass of the bound state belongs to the interval $(m_H,2m_H)$, thus the lowest limit approaches the mass of the Higgs. In this respect, this method allows to study situations in which the mass of the bound state is also well below the threshold, thus represents an improvement w.r.t. the on-shell scheme. Nevertheless, the results should be comparable for what concerns the emergence of the bound state. This turns out to be the case, as we will show in the next section.\par

\section{Results and Discussion\label{sec:RnD}}
In this section, we present the results of the present study. To this end, the tree-level amplitude obtained in Eq. \ref{eq:scatamp} was unitarized using the two schemes discussed in Sec. \ref{sec:unit}. 
The unitarized amplitudes provide bounds for the values of the coupling constants that do not allow for a bound state, as summarized in Fig. \ref{fig:bounds}.\par

Namely, let us first start with the on-shell scheme by assuming (in agreement with present experimental data) that no Higgsonium exists, in the on-shell scheme the corresponding parameter space in the $({d_3,d_4})$-plane lies below the dot-dashed blue line shown in Fig. \ref{fig:bounds}. This curve corresponds to the equation
\begin{align}
d_4^c &= 5 d_3^2-\frac{64\pi v^2}{\sqrt{3} m_H^2} \text{ ,}
\end{align}
which can be (also) obtained from the divergence of the scattering length. Furthermore, the following condition holds true. For a fixed $d_3$,
\begin{align}
    \Bigg\{&\begin{matrix} d_4>d_4^c & \text{no bound state}\\
     d_4<d_4^c & \text{bound state}\end{matrix}
     \text{ .}
\end{align}

Next, the $N/D$ unitarization scheme delivers qualitatively similar results as can be also seen from Fig. \ref{fig:bounds}. The dashed red curve in Fig. \ref{fig:bounds} represents the upper bound for the coupling constants obtained from the $1^{st}$ iteration of the $N/D$ scheme ({\it i.e.,} $D^{(1)}(s)$). 
Successive iterations do not alter this curve by any large amount, as discussed in detail in Appendix \ref{appendix}.\par

For completeness, we also discuss the bounds to the coupling constants $d_3$ and $d_4$ given by the demand that $h=0$ corresponds to the global minimum of the potential. (These results have been presented in, {\it e.g.,} Ref. \cite{Agrawal:2019bpm}; we briefly recall them here for the sake of clarity.)
Referring to Eq. \ref{sympot3}, the extrema of the potential are given by the zeroes of the first derivative of the potential with respect to the field. Considering only the first three terms,
\begin{align}
\frac{dV(h)}{dh} &= m_H^2 h + \frac{g}{2}h^2 + \frac{\lambda}{3!} h^3\nonumber\\
   & = m_H^2h + d_3 \frac{3m_H^2}{2v} h^2 + d_4 \frac{3m_H^2}{6v^2}h^3.
\end{align}
Thus, the extrema of the potential are,
\begin{align}
h=0, \text{ and } & h = \frac{v}{2d_4}\left(-3d_3 \pm \sqrt{9d_3^2-8d_4}\right).\label{eq:Higgsterma}
\end{align}
The first root ($h=0$) corresponds to the EW vacuum (more precisely, the point $H=v$), and is a local minimum (as can be verified from the second derivative of the potential). The other two roots form a pair of minimum and maximum of the potential. In the SM ($d_3=1=d_4$) these points reduce to $h=-2v$ and $h=-v$ respectively (equivalently, $H=-v$ and $H=0$). These two roots are real only if $d_3 \ge \sqrt{\frac{8d_4}{9}}$. Thus, if we demand that the root $h=0$ represents a global minimum, we get the lower bound for the parameters \cite{Agrawal:2019bpm}.
The corresponding curve is represented by the solid black line in Fig. \ref{fig:bounds}. It should be also stressed that the above derivation does not require that the potential is $Z_2$-symmetric around $h=-v$. Note that, the curves, including the on-shell and $N/D$, are symmetric under $d_3\to-d_3$. This is due to the fact that the tree-level amplitude is quadratic in $d_3$. But, this is to be expected as $d_3$ characterises the attractive part of the Higgs potential and a change in the sign will flip the positions of the maxima and the minima given in Eq. \ref{eq:Higgsterma}.\par

The bounds on the value of $d_3$ obtained here lie close to the ones measured by the experiments \cite{ATLAS:2019qdc,ATLAS:2022jtk,CMS:2022dwd}. Particularly, the value obtained from the double-Higgs pair production processes is consistent with the lower limit obtained in the present work \cite{ATLAS:2019qdc}. However, additional data and the inclusion of the single Higgs production measurements tightens the experimental bounds \cite{ATLAS:2022jtk}, which has also been corroborated by the resonant production channel measurements \cite{CMS:2022dwd}. 
In this case, we conclude from our work that the Higgsonium cannot exist, in the SM as well as beyond it. However, it would
serve useful to continue the search for Higgs-Higgs self interactions, as it can provide rich
information about the SM and eventual extensions of it.\par

Summarizing, it is visible from Fig. \ref{fig:bounds} that the non-existence of the Higgsonium further constrains the admissible region of the $({d_3,d_4})$-plane and constitutes a more stringent constraint w.r.t. the global minimum one of Ref. \cite{Agrawal:2019bpm} in the upper part of the parameter plane.\par

Next, we present other quantities relevant quantities for the the Higgs-Higgs scattering. Using the on-shell scheme, we plot the scattering phase shift in Fig. \ref{fig:phshift}. The SM case is basically not distinguishable from the horizontal axis, implying that the phase-shift is always very small (in agreement with the fact that the tree-level result offers a very good approximation). 
In general, the phase-shift is positive when no Higgsonium is present, but is negative otherwise. Note, in the latter case the $S$-wave scattering phase shift goes to $-2\pi$ as $s$ increases, thereby confirming the presence of a bound state \cite{Ma:2006zzc} (in fact, two states are present below threshold: a bound state and the Higgs particle, each of them generating a $-\pi$ shift). \par

In Fig. \ref{fig:scatlen} we plot, for the $N/D$ case, the scattering length as a function of the 4HC coupling $d_4$ for two illustrative values of $d_3$. For $d_3=d_4 =1$ (SM value, left plot), the scattering length is positive; then, for increasing $d_4$ it decreases and eventually becomes negative. The smooth behavior of the scattering length indicates the absence of any bound state\footnote{It must be noted that, the unitarization methods considered here are applicable only when the bound state is close to the 2-particle threshold. Hence, the self-duality problem cannot be explored using this method.}. 
For the illustrative large value of $d_3=15$ (right plot), the scattering length is negative for any value of $d_4\lesssim 420$ signalling the presence of the Higgsonium. The divergence at $d_4 \sim 420$ implies that the bound state is realized at threshold. For larger values of $d_4$ there is no bound state.\par

Finally, we show in Fig. \ref{fig:2hmass} the mass of the Higgsonium as derived from the pole of the unitarized amplitudes for a fixed value of $d_4=272$, as a function of $d_3$. As the strength of attractive interactions ($d_3$) is increased keeping the repulsion ($d_4$) fixed, the depth of the bound state increases. The converse is true for $d_4$, the depth of the bound state for a given $d_3$ decreases as repulsion is increased. 
In the end, it should also be noted that, by construction, the mass of the Higgsonium goes to $3m_H^2$ ($m_H^2$) in the on-shell ($N/D$) unitarization scheme as $d_3\to\infty$, implying that an agreement is only possible when the bound state is not too deep. Yet, the results are quite similar for all the range $d_3 \lesssim 20$ under investigation.
\begin{figure*}[ht]
\centering
\includegraphics[scale=0.65]{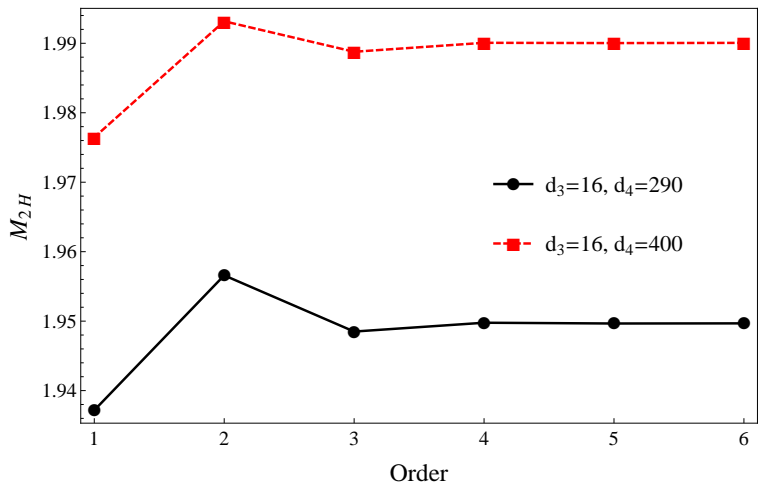}
\caption{Plot showing the mass of the Higgsonium at different orders of the $N/D$ unitarization. The vertical axis is in the units of $m_H$, see text in Appendix \ref{appendix} for details.}\label{fig:2hmassorder}
\end{figure*}

\section{Summary and Conclusions\label{sec:SnC}}
In this work, we have studied the scattering of two Higgs bosons. The tree-level scattering amplitudes are derived from the generalised Higgs potential and the scattering lengths for the S-, D-, and G-waves have been evaluated (the particular case of the SM parameter has been also outlined). 

Then, the scattering amplitudes are unitarized using the well-known on-shell and the $N/D$ unitarization schemes. The SM scattering lengths are basically left unchanged, thus offering a clear prediction for the SM outcomes. 
Away from the SM values, the unitarization renders the investigation of an eventual bound state possible. Both procedures indicate the presence of a bound state of two Higgs bosons (Higgsonium) provided the the 3H self-interactions of the Higgs boson is at least $10$ times larger than the value predicted by the SM. Indeed, the requirement that the Higgsonium does {\it not} exist generates some constraints for the 3H and 4H coupling that are more stringent than the ones obtained by the global minimum requirement in the upper part of the 3H-4H coupling plane, and in line with the current experimental bounds for $d_3$ within the uncertainty levels.\par

In the future, experimental searches for Higgs-Higgs scattering (and, in a broader sense, for a Higgsonium) could turn up some interesting results and help us further constrain the SM results. It is also possible (although at present quite speculative) that a two-Higgs resonance state exists well above threshold \cite{Consoli:2020kip} and the constraints one can derive from the study of such a state could be used in conjunction with those from the present work to further narrow down the 3HC and 4HC parameters.\par

\section*{Acknowledgements}
F.G. thanks A. Pilloni and E. Trotti for useful discussions. The authors thank the referee for bring to our attention Refs. \cite{ATLAS:2022vkf,ATLAS:2022jtk,CMS:2022dwd}. 
The authors acknowledge financial support from the Polish National Science Centre (NCN) via the OPUS project 2019/33/B/ST2/00613.\par\medskip

\appendix
\section{$N/D$ at higher orders\label{appendix}}
In this appendix, we explore the effects of going to higher orders in the $N/D$ unitarization scheme. The variation of the mass of the bound state as a function of the order ($k$) of the denominator ($D^{(k)}(s)$) is shown in Fig. \ref{fig:2hmassorder}. A slight variation is seen in the mass when going from $k=1$ to $k=2$ and $k=3$ in the numerical evaluation of the $N/D$-unitarized amplitude. However, the value stabilizes soon after and for $k\ge 4$ the mass of the bound state is fixed.\par
Plugging Eq. \ref{eq:Nk} and \ref{eq:Dkplus1} into each other iteratively, one finds that the denominator at order $k$ can be written as,
\begin{align}
    D^{(k)}(s) &= D^{(1)}(s) + I(s) - C^{(k-1)}(s)\text{;}~~~~~k\ge 1 \text{ ,}\label{eq:denomND}
\end{align}
where, 
\begin{align}
    C^{(k-1)}(s) &= \frac{(s-m_H^2)}{\pi^2} \int_{4m_H^2}^{\infty} ds_1 \int^{3m_H^2}_{-\infty} ds_2 \nonumber\\
    &\frac{\sigma(s_2)\rho(s_1)D^{(k-1)}(s_2)}{(s_2-s_1-i\epsilon)(s_1-s-i\epsilon)(s_1-m_H^2)} \text{ ,}\label{eq:intC}\\
    I(s) &= \frac{(s-m_H^2)}{\pi^2} \int_{4m_H^2}^{\infty} ds_1 \int^{3m_H^2}_{-\infty} ds_2\nonumber\\
    &\frac{\sigma(s_2)\rho(s_1)}{(s_2-s_1-i\epsilon)(s_1-s-i\epsilon)(s_1-m_H^2)} \text{ .}\label{eq:intI}
\end{align}

Above, $D^{(1)}(s)$ is as given in Eq. \ref{eq:D1approx}. In Eq. \ref{eq:denomND}, the first two terms are independent of the order of approximation of the denominator (note, $I(s)=C^{(0)}(s)$). In the integrand of $I(s)$, $0<\rho(s_1)\le 1/32\pi$, $|\sigma(s_2)|\le 2g^2/m_H^2$. Also, $|(s_2-s_1-i\epsilon)(s_1-m_H^2)|>3m_H^4$ owing to the domains of integrations. Thus, we can impose an upper bound for the integral given in Eq. \ref{eq:intI} as,
\begin{align}
    |I(s)|&< \frac{(s-m_H^2)}{\pi^2}\frac{g^2}{48m_H^6} \Bigg|\int_{4m_H^2}^\infty \frac{ds_1}{(s_1-s-i\epsilon)}\Bigg|\text{ .}
\end{align}
Using the same arguments, we can show from Eq. \ref{eq:D1approx} that, $|D^{(1)}(s)|<1$. 
These bounds can then be used to show that $|D^{(k)}-1|\ll 1$ as long as the tree level amplitude is finite in the region around the expected pole. It then follows that the contribution of $C^{(k-1)}(s)$ is small in the vicinity of $s=s_{pole}$, provided $s_{pole}\in(3m_H^2,4m_H^2-\delta)$.
Thus, close to the pole of the possible bound state, as the order of the denominator increases the corrections keep reducing in magnitude. This behavior carries over to the position of the zero of the denominator $D(s)$, as demonstrated in Fig. \ref{fig:2hmassorder}. This also means that the upper bound imposed on the value of the 4HC parameter $d_4$ does not vary significantly from that derived from the $1^{st}$ iteration (shown in Fig. \ref{fig:bounds}).

\end{document}